\documentclass[11pt]{article}
\pdfoutput=1 
\usepackage{pdfpages}
\usepackage[margin=1in]{geometry}

\title{Research Pearl: The ROSI Operating System Interface}
\author{Robert Soul\'{e} \and Peter Alvaro}
\date{}

\begin{document}

\maketitle

Korth and Silberschatz published the following paper as a technical
report from the University of Texas at Austin in 1984. We have chosen
to republish the report on arXiv for two reasons. First, it represents
an important piece of early work that examines the relationship
between operating systems and databases. Second, the original
publication is not readily available online.

The relationship between operating systems and databases has always been complex.
In their earliest days, databases presented an
ideologically-distinct alternative to general-purpose operating systems
 for solving the problems of resource
and data managment~\cite{notes}.
For the past four decades,  database have been designed as
applications that run
on top of an operating system, making use of the operating system's
facilities for managing hardware resources.  However, databases and
operating systems are often at odds on how to manage those
resources.  Cache replacement, scheduling, and file management are just
a few examples of low-level functionality that is typically redundantly implemented
in the database engine to take advantage of application-level semantics that
are opaque to the OS~\cite{stonebraker94}.


Korth and Silberschatz's ROSI is the first work that we are aware of
to propose a more extreme position: that we should treat an operating
system as if it were a database. Their motivation is usability. They
argue for replacing traditional file systems with a relational data
model, asserting that it would provide a uniform interface that would
be easier to learn and allow for greater portability. The paper
includes several examples of what a relational interface to an
operating system might look like, including SQL queries for searching
for files, examining I/O request queues, and listing user
information. The second half of the paper is more theoretical,
discussing how the universal relation model can be applied to
operating systems.

In recent years, we have seen renewed interest in inverting the
traditional operating-system and database relationship. For example
Boom~\cite{Alvaro10} proposes a data-centric approach to cloud
management, and DBOS~\cite{Skiadopoulos21} have argued for a
clean-slate cloud operating system implemented as a database.
Korth and Silberschatz's technical report shows the intellectual 
history of many of these contemporary systems.

{\tiny
\bibliographystyle{abbrv}
\bibliography{main}
}



\section*{Supplementary material}

ROSI: A USER-FRIENDLY OPERATING
SYSTEM INTERFACE BASED
ON THE RELATIONAL DATA MODEL

Henry F. Korth
Abraham Silberschatz

Department of Computer Sciences
University of Texas at Austin
Austin, Texas 78712

## Abstract

This paper presents some preliminary results concerning a new user-friendly operating system interface based on the relational data model that is currently under development at the University of Texas at Austin. The premise of our work is that a relational model of the operating system environment will produce a user and programmer interface to the system that:

- is easier to use
- is easier to learn
- allows greater portability

as compared with existing operating system interfaces. Our approach is to model elements of the operating system environment as relations and to model operating system commands as statements in a relational language.

In adapting the relational model to an operating system environment, we found it necessary to extend the model and improve existing relational languages. The extensions to the relational model are designed to allow a more natural representation of elements of the environment. Our language extensions exploit the universal relation model and utilize the graphical capabilities of modern workstations. The nature of our investigations is ranging from practical implementation issues to the more theoretical questions of modeling and language semantics.

# 1. Introduction

File management is one of the most visible services of an operating system. A file is a collection of related information defined by its creator. Commonly, files represent programs (both source and object forms) and data. Data files may be numeric, alphabetic or alphanumeric. Files may be free-form, such as text files, or may be rigidly formatted. In general, a file is a sequence of bits, bytes, lines or records whose meaning is defined by its creator and user.

Computers can store information in several different physical forms; magnetic tape, disk, and drum are the most common forms. Each of these devices has its own characteristics and physical organization. For convenient use of the computer system, the operating system provides a uniform logical view of information storage. The operating system abstracts from the physical properties of its storage devices to define a logical storage unit, the *file*. Files are mapped, by the operating system, onto physical devices.

Many different types of information may be stored in a file: source programs, object programs, numeric data, text, payroll records, and so on. A file has a certain defined *structure* according to its use. A text file is a sequence of characters organized into lines (and possibly pages); a source file is a sequence of subroutines and functions, each of which is further organized as declarations followed by executable statements; an object file is a sequence of words organized into loader record blocks.

A major consideration is how much of this structure should be known and supported by the operating system. If an operating system knows the structure of a file, it can then take advantage of this structure in operations on the file. For example when a user tries to print the binary object form of a program, the attempt normally produces garbage. This can be easily prevented *if* the operating system has been told that the file is a binary object program.

There are disadvantages to having the operating system know the structure of a file. One problem is the resulting size of the operating system. If the operating system defines many different file structures, it must then contain the code to support these file structures correctly. In addition, every file must be definable as one of the file types supported by the operating system. Severe problems may result from new applications that require information structured in ways not supported by the operating system.

For example, assume that a system supports two types of files: text files (composed of ASCII characters) and executable binary files. Now if we (as a user) want to define an encrypted file to protect our files from being read by unauthorized people, we may find neither file type to be appropriate. The encrypted file is not ASCII text but (apparently) random bits. But though it may appear to be a binary file, it is not executable. As a result, we may have to circumvent or misuse the operating system's file types mechanism, or modify or abandon our encryption scheme.

The other extreme is to impose (and support) no file type in the operating system. This approach has been adopted in UNIX, among others. UNIX considers each file to be a sequence of 8-bit bytes; no interpretation of these bytes is made by the operating system. This scheme provides maximum flexibility, but minimal support. Each application program must include its own code to interpret an input file into the appropriate structure.

Special file structures may be provided in different ways in different operating systems. Thus, it may be difficult to design applications that run under several operating systems. To alleviate this problem, it is desirable that file types be defined in a high-level, implementation-independent manner.

We believe that the operating system should provide the user with a single higher-level conceptual view of data, rather than merely a collection of files. This approach has two major advantages:

(1) The operating system needs to deal with only one construct which makes implementation and maintenance simpler.

(2) The operating system interface becomes more uniform, and as a result, easier to master. Since many of the features of the operating system command language deal with file systems (physical resources in many systems are treated just like regular files), such an approach allows the user to deal with a uniform, well-understood interface.

The higher level view we are advocating is based on the *relational data model*. We thus propose replacing the conventional file system traditionally supported by the operating system with a relational system. The premise is that a relational model of the operating system environment will produce a user and programmer interface to the system that:

- is easier to use
- is easier to learn
- allows greater portability

as compared with existing operating system interfaces. Our approach is to model elements of the operating system environment as relations and to model operating system commands as statements in a relational language.

In adapting the relational model to an operating system environment, we found it necessary to extend the model and improve existing relational languages. The extensions to the relational model are designed to allow a more natural representation of elements of the environment. Our language extensions exploit the universal relation model of Fagin, Mendelzon, and Ullman [1982] and utilize the graphical capabilities of modern workstations. The nature of our investigations is ranging from practical implementation issues to the more theoretical questions of modeling and language semantics.

## 2. Motivation

In the relational data model, the database consists of a collection of tables, called *relations*. Each relation $R_i$ is described by a *relation scheme* $R_i$ ($A_1$, ..., $A_n$), where each $A_j$ denotes an *attribute* of $R_i$. If we view the relation as being a table, the attributes correspond to column headers. Each row of the table is a *tuple* of the relation.

A relational database system includes a set of well-defined operations to access and modify the data. There are several relational data languages. For simplicity, we shall restrict attention to SQL (Chamberlin, et al. [1976]).

The relational data model is considered the easiest to use of the three primary data models (network, hierarchical, and relational). The main reason is that the relational model provides a conceptual view of the data that is unencumbered with details of the underlying implementation. Using the relational model, it is possible to query the database using a nonprocedural language (i.e., the user describes what is wanted, rather than providing a method to find it).

To illustrate our ideas, we present several examples of our approach to the modeling of the operating system environment. Specifically, we present several examples of relations which one may expect to encounter in an operating system environment. We elaborate on our plans to create a working version of this model in subsequent sections.

### 2.1. Mail System

One function of an operating system is to allow users to communicate with each other by exchanging messages. Such a mail system could be represented in our scheme by the two relations:

    in-mail (sender, cc-list, subject, date-received, text)

    out-mail (to, cc-list, subject, date-sent, text)

Suppose that we wish to list all the messages that have been received from John Doe. Using, for example, the query language SQL, we write the following:

```
select *
from in-mail
where sender = "John Doe".
```

Suppose that a user wishes to send a message to user John Jones. This can be done by adding the tuple:

("John Jones", "next meeting", , ...)

to the out-mail relation. The system will add the appropriate date to the tuple. System-provided values for fields of an update is a feature not normally provided in database systems.

Notice that in this example, the mail system is treated just like a relation. Thus a user familiar with the command language associated with the relational operating system environment need not learn a different command language associated with the mail system. In contrast, current operating systems offer one or more mail programs. These mail programs are not considered to be part of the operating system, per se. Rather they are applications that are free to provide their own user interfaces, independent of the rest of the environment.

### 2.2. Text Files

Consider an ordinary text file that contains a technical paper authored by Hank Korth and Avi Silberschatz, titled "A tutorial on graph-based locking protocols," and submitted to the *ACM Computing Surveys* on October 12, 1983. Using a traditional file system, one *cannot* record the information about author, title, journal submitted to, and date submitted, with the file itself. Rather, one must create a new text file and record this descriptive information as arbitrary text. The user is responsible for maintaining the connection between this descriptive information and the text itself. This is both cumbersome and complicated. Using our approach, one could define a text file as the following relation:

technical-papers = (title, authors, journal, submission-date, text)

Suppose that we wish to list all the papers that were submitted to *ACM Computing Surveys*. Using, again, the query language SQL, we write the following:

```
select title
from technical-papers
where journal = "ACM Computing Surveys"
```

Notice that in this example, all the needed information is concentrated in a single place and that our command interface can be effectively used in issuing the described queries.

Document retrieval capability of this sort is not part of the directory management component of any operating system we are aware of. Rather, certain systems, typically office-automation systems, include a document-retrieval application that runs on top of the file system.

Note that we have included the text of the paper as an attribute of the relation. We do so because the text is logically a single object at the directory level. We shall see more clearly how text can be represented in Section 3.

### 2.3. Request Queues

An operating system must maintain a queue of pending I/O requests to the various devices supported by the system. Such an I/O queue can be maintained as the following relation:

queue (file-name, user-name, size)

User Avi, wishing to examine the status of his I/O request, can do so as follows:

```
select *
from queue
where user-name = "Avi"
```

User Avi, wishing to print file A, can do so by adding the tuple (A, , ) to the queue relation. The system, in turn, will add the user name and file size to the tuple. Notice that the "print A" command commonly available in any system is simply a syntactic sugar for adding the appropriate tuple to the relation queue.

A user may have several requests pending (printing, mail being delivered, ...). A pending-request relation:

pending (job-name, request #, request-type, status)

can be maintained. The user may query this relation to check the status of tasks being performed asynchronously by the operating system on the user's behalf.

### 2.4. User Information

Many operating systems maintain a summary of information about the various users of a system. This summary may be viewed by the user community. In addition, a user is allowed to modify the information about himself or herself. Such information may be maintained in a relation with the following attributes:

user-info (user name, mail-address, job, subsystem, TTY, logged-in, ...)

A user wishing to obtain the summary about all the users currently logged in can do so by:

```
select *
from user-info
where logged-in = "on"
```

We will later show that this query can be made more concise using views.

### 2.5. Summary

In current systems, each of the elements of the environment we have discussed is managed by a separate application with its own interface. We have seen that a single model may encompass all of them and a common language may be used to manipulate them. The user need not remember several sets of often cryptic commands and option codes. The user *is* required to learn a relational language. We address the ease-of-learning question in Section 4.

The support for text files discussed above can be extended to support all of the file-related information stored in a typical operating system's directory (e.g., file-name, file type, location, creation-date, date of last access, protection). We shall elaborate on directory management in the next section.

The relational model does more than simplify the user interface. It also enhances its power. Typically, a user can list part of a directory using a regular expression to select on file-name. The relational interface allows selection based on *any* set of fields in the relation.

### 3. Relational File System

In this section, we outline some of the topics that we are in the process of investigating so that one can replace a file system with a relational system.

### 3.1. Generic Relation Types

A relation scheme may be re-used for several relations. For example, each user has an *in-mail* relation. Associated with each attribute in a relation scheme is a *domain*, representing the set of permissible values for each column of the table. Some of the types are predefined by the system, while others are user defined. When a new relation is created, its type must already be defined in the system. Thus for each type several instances may coexist in the

system.

Examples of system defined types are the in-mail, out-mail relation schemes described in Section 2.1, relation schemes for text files, program files, appointment calendars, etc. Examples of user defined types are relation schemes for recording grades in a class (e.g., a relation with attributes (student-name, midterm-grade, final grade)), or for keeping track of bank balances.

Given the concept of a relation type, we need to provide a mechanism to allow type augmentation. That is, a user should be allowed to create an instance of type with fewer attributes or additional attributes. The idea of creating a type with fewer attributes has been previously discussed in the literature in the context of *views* (see e.g., Keller [1981]). Some database systems that support variable-length records allow adding attributes. However, little work has been done on the problem of adding new attributes in the construction of a view to a base relation. For example, a user may wish to augment his in-mail relation with a new attribute, "status," with domain = {urgent, deferred, personal, junk}.

Adding new attributes results in complications in the underlying implementation as well as in the semantics of insert and update. Users referencing the base relation are now referencing a view of the augmented relation. Nevertheless, it is a feature that may enhance significantly the naturalness of the user interface.

To achieve this, we borrow the concept of view from database theory. Thus we can define a view "finger" by the SQL statement:

    **define view** finger **as select** user-name, job, TTY
                      **from** user-info
                      **where** logged-in = "on"

Using this view, the example of Section 2.4 can be simplified to

    **select** *
    **from** finger

## 3.2. Directory Structure

A central part of a file system is the directory structure, whose main purpose is to keep the necessary information about each file in the system (e.g., file-name, file type, contents, size, protection).

Most directory structures are organized as trees or directed graphs. If we wish to preserve these types of structures in our relational system, we need to allow a system to define domains of type relation. Many systems incorporate such a feature in the *system catalog*. The system catalog is a relation with attributes relation-name, number-of-attributes, size, etc. In order to allow the modeling of graph structured file systems, the system-catalog notion must be made recursive. While existing systems frequently treat the system catalog as a relation, the notion of relations within a relation that the system catalog represents is treated as a special case.

To allow this capability in its full generality, we provide the following generic relation type:

    directory = (relation-name, type, access, content)

The domain of attribute "type" includes system-defined types as well as user-defined types. The domain of attribute "content" includes all possible instances of all the supported generic types. Thus for example, if we have a tuple:

    (project, directory, RW ---, ...)

then the last field in the tuple represents an instance of a relation of type *directory* that may be as shown in Figure 1. Each entry in the column labeled "content" consists of the actual source code written in some programming language. We note that the "type" attribute may be used in selecting the appropriate special purpose program for operating on the

| relation-name | type | access | content |
|---|---|---|---|
| Program 1 | Pascal | RW --- | ... |
| Program 2 | Lisp | E --- | ... |
| . | | | |
| . | | | |
| . | | | |
| Program 9 | Prolog | E --- | ... |

Figure 1: Relation of Type Directory

"content" field. For example, if the type is text, then a text editor is selected. Similarly, if the type is a directory, then the content field may be operated upon using our relational language.

An instance of a directory type thus includes information about all the relations in one particular subdirectory. For example, the root directory of a particular user may look as shown in Figure 2.

The relation Book represents a subdirectory consisting of the various chapters of a book being written by the user. The user may wish to augment the base relation directory with the title of the chapter resulting in an augmented relation type book-directory with attributes: (relation-name, type, access, title, content). Thus the book subdirectory may be represented as in Figure 3, where the last field in each tuple contains the actual text of each chapter in the book.

Thus, in general, one can construct an arbitrary hierarchy of relations. For the example above, the hierarchy is depicted as in Figure 4.

| relation-name | type | access | content |
|---|---|---|---|
| Project 1 | Directory | RW --- | ... |
| Paper 1 | Technical-paper | RW --- | ... |
| My-Mail | In-mail | R --- | ... |
| . | | | |
| . | | | |
| . | | | |
| Book | Book-Directory | RW --- | ... |

Figure 2: A Root Directory

| relation-name | type | access | title | content |
|---|---|---|---|---|
| Chapter 1 | Text | RW --- | Introduction | ... |
| . | | | | |
| . | | | | |
| . | | | | |
| Chapter 8 | Text | RW --- | The Deadlock Problem | ... |
| . | | | | |
| . | | | | |
| . | | | | |

Figure 3: A Subdirectory Structure

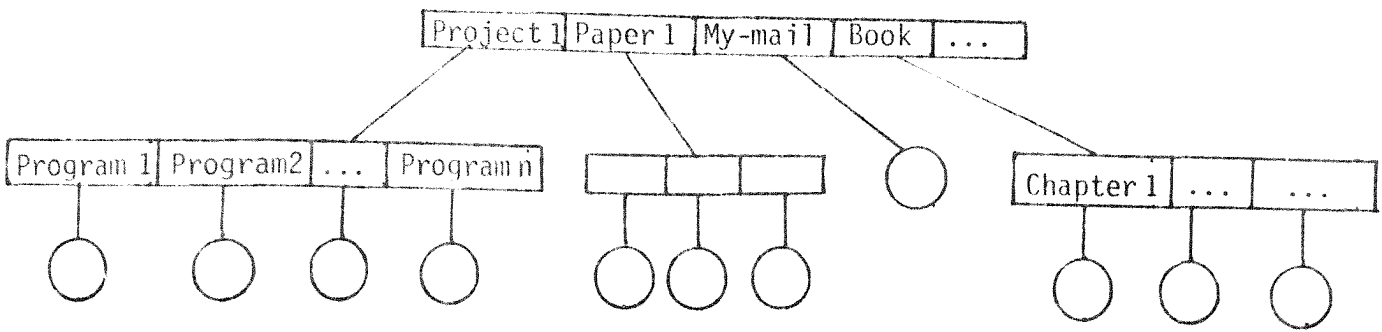

Figure 4: A Hierarchical Directory Structure

### 3.3. Naming, and the Structure of the User's Environment

In current file systems, hierarchical or otherwise, files are referenced using the file-name. In current operating systems, no other identifier is available for selection of files. In our relational file system model, selection is available by any attribute or set of attributes in the directory relation. Thus files may be referenced by title, date, etc.

The above observation suggests adding a *keyword* field to the directory relation. We may now associate keywords with files in the same way that document-management systems allow keywords for documents. This allows us to query the file relation to find files with a desired set of keywords. For example:

**select** *
**from** directory
**where** {dinosaur, compiler} ⊆ keywords

The response to this query is, in general, a set of files, rather than a single file. Rather than resorting to file-names to identify uniquely the desired file, the user may simply place the cursor next to the entry for the desired file. We elaborate on this in our discussion of graphical queries in Section 5.5.

In many existing operating systems, file names may be selected based upon a *regular expression*. For example, **comp.*er** denotes all strings beginning with "comp" and ending in "er". We are now in the process of extending SQL to incorporate this feature.

The keyword approach has significant implications on the style of file system we may see in future operating systems. Many users may choose to have one directory relation rather than a hierarchy of relations. Such users will rely exclusively on keyword-based retrieval to access files rather than using the user-guided tree traversal of operating systems such as UNIX.

Listing the contents of a directory requires little work by a traditional system. The directory entries typically are stored contiguously on disk. Thus one disk access suffices for the retrieval. A keyword retrieval requires more disk accesses. If there is no index, all files belonging to the user must be examined. For users with many files this is unacceptable. An efficient keyword-based retrieval scheme is presented in King, Korth, and Willner [1983]. This scheme reduces the number of disk accesses required, but it is still higher in overhead than a simple hierarchical system. We note, however, that the efficiency advantage of hierarchies is lost if extensive use is made of links to represent a general graph structure.

The payback of the added overhead is a gain in ease-of-use. As personal workstations become more prevalent, the added overhead will become less significant, provided an efficient index scheme is used.

Observe that system commands are, in fact, nothing more than executable programs. Thus, they may be represented as files in the user view. Therefore, users who are not skilled system users may find commands using the same query system used to find data files. A result of this approach is a symmetry between commands and the objects they operate on. The descriptive information stored in the system about the operands to a command can be accessed by the code implementing the command. Thus, for example, the print command can determine the type of file to be printed and direct it to the appropriate formatter and/or printer without user intervention. It can detect errors such as the case we referred to in the introduction, of attempting to print a file containing the binary object code of a program.

Because of the object/command symmetry, we can also query the system as to which commands are applicable to a given file.

This sort of information can support an object-oriented user interface in the style of *Smalltalk* (Goldberg and Robson [1983]), and *Smallworld* (Hailpern and Laff [1983]). We are not proposing to construct an object-oriented system. We mention it only to illustrate that the relational operating system environment provides an excellent foundation for the construction of easy-to-use user interfaces.

## 4. A Simplified User-View of the Database

The goal of our discussions above has been to simplify the conceptual model of a file system by adding semantic information and a query language that allows reference to this information. However, this added power comes at a price. Simple tasks have become more complex as we have simplified complex tasks.

To illustrate this, suppose the user has a file called "memo" that is to be printed. Instead of saying "print memo", the user says:

insert &lt;memo, ...&gt; into print queue.

Thus, we need to provide an easy-to-use interface that allows users to perform simple operations more easily. There are several types of simplification to the user interface.

- *synonyms* -- We may represent a commonly used database operation by a shorthand.
- *simplified user-view of the database* -- Recent research in universal-relation databases promises enhancements to the ease-of-use of the relational database model.
- *two-dimensionality* -- Commands and queries may be expressed in ways that exploit the two-dimensional nature of workstation displays. The addition of color provides increased power to simplify the user interface.

The first item above is self-explanatory. We now elaborate on the other two.

### 4.1. Universal Relation Model

The approach we are taking to simplifying the user view is based on the *universal relation model* (Ullman [1982]). In this model, the user sees only one relation (the universal relation). The system determines which relations of the underlying database need to be accessed in order to respond to a given query. The semantics of the database are expressed using a *hypergraph* in which the nodes are the attributes of the universal relation, and the hyperedges are fundamental relationships (called *objects*). A second hypergraph is then formed whose nodes are the objects, and whose hyperedges, called *maximal objects* represent maximal sets of objects in which queries "make sense."

The result of research in this area is query languages that are similar to SQL, but lacking the from clause. Examples are System/U (Korth, et al. [1984]) and Pique (Maier et al. [1981]).

The remainder of this section describes the universal relation model. In Section 5, we show how we can apply this model to our relational operating system environment. We

assume familiarity with the relational data model as covered in a text such as Ullman [1982]. In particular, we assume familiarity with the notions of functional dependencies (FDs), multivalued dependencies (MVDs), and join dependencies (JDs).

To illustrate the universal relation concept, we present a simplified example of banks, accounts, and customers with U (bank, account, customer).

The universal relation is defined as the set of all tuples such that some predicate is true:

$$\{ t \mid P(t) \}$$

The predicate $P$ must pertain to individual tuples if the database is to be a relation. In the bank example, $P$ would be "$t$.customer has $t$.account and $t$.account is with $t$.bank." Let us agree to write $P(t)$ in conjunctive normal form:

$$P(t) = P_1(t.a_1^1,...,t.a_{n_1}^1) \text{ and } \cdots \text{ and } P_k(t.a_1^k,...,t.a_{n_k}^k)$$

where $t.a_j^l$ denotes an attribute of $t$. Let us agree further to treat each set of attributes $\{ t.a_1^l,...,t.a_{n_j}^l \}$ as a fundamental relationship, or object. The predicate $P$ may now be written as

$$P(t) = P_1(O_1) \text{ and } \cdots \text{ and } P_k(O_k)$$

where the $O_i$'s are the objects.

In the bank example the objects are:

- $O_1 = $ (bank, account)
- $O_2 = $ (account, customer)

The above appears to be a highly intuitive definition of the semantics of the universal relation. However, this definition turns out to have significant theoretical implications. Fagin, Mendelzon, and Ullman [1982] showed that for any universal relation $U$ defined as we have defined it above, the join dependency $|\times|(O_1,...,O_k)$ holds. Furthermore, they characterized the MVDs that the above JD imply. Their characterization is best understood by considering the database to be a hypergraph whose nodes are the attributes of $U$ and whose hyperedges are the objects. A MVD of the form $X \rightarrow\rightarrow Y$ (where $X$ and $Y$ are disjoint) holds on $U$ if and only if $Y$ is the union of connected components of the hypergraph with nodes in $X$ deleted. The hypergraph for the bank database appears in Figure 5.

It is not practical to implement a database system by storing the universal relation itself. Rather, one decomposes $U$ into several relations. However, if we decompose $U$, we are faced with the problem of translating queries on $U$ into queries on the underlying database system. It is often the case that objects correspond to relations in the underlying database (though it suffices that objects be the projection of relations). Thus, translating a query on $U$ to one on the actual relations amounts to finding a path in the hypergraph that includes at least all of the attributes mentioned in the query.

In the bank example, the proper path in the hypergraph was obvious. Consider a more complicated banking example in which we also have loans as well as accounts. The hypergraph for this expanded example appears in Figure 6. Now the query

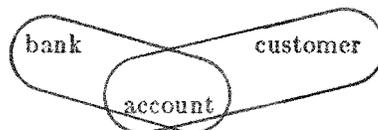

Figure 5: Hypergraph Showing the Objects for the Bank Database.

```
select customer
where bank = "InterFirst"
```
has several possible interpretations:

- Find all customers with an account at InterFirst.
- Find all customers with a loan at InterFirst.
- Find all customers with both an account and a loan at Interfirst.
- Find all customers who do business with InterFirst (i.e., have *either* a loan or an account at InterFirst).

These interpretations arise from the fact that there are two paths in the hypergraph from bank to customer. We can take one or the other, or we can take both paths and then take either the union or intersection of the results.

It is not our point here to argue which interpretation is correct. Rather, we wish to illustrate how one can alter the semantics of the universal relation to specify which interpretation is desired. Maier and Ullman [1983] define another hypergraph whose nodes are the objects. The hyperedges of this hypergraph, called *maximal objects*, are the largest collections of objects in which we are willing to let the system navigate automatically. Assuming the fourth interpretation for the bank database, the maximal objects would be M1 and M2 where:

M1 consists of objects (bank, account), and (account, customer)
M2 consists of objects (bank, loan), and (loan, customer).

A query is processed in all maximal objects that contain all of the attributes mentioned in the query. The union of all responses thus obtained is the answer to the query. Queries are processed within maximal objects by taking the join of all objects in the maximal objects.

A heuristic for constructing maximal objects directly from the data dependencies appears in Korth, et al. [1984]. A critical part of this heuristic is the notion of *cycles* in hypergraphs. Intuitively, a cycle represents a potential ambiguity in query interpretation. We normally wish to construct maximal objects whose member objects represent a acyclic subhypergraph.

System/U is a database system based on this data model. We are currently using System/U to create an easy-to-use relational model for our operating system environment.

### 4.2. Graphical Queries on the Universal Relation

The System/U query language is a one-dimensional language, as is SQL. To enhance ease of use we are developing a two-dimensional, graphical relational language that exploits the power of bit-mapped workstations and pointing devices (mouse, tablet, light-pen, etc.). The design of our language involves two areas of research:

(1) *Hypergraph representation theory* -- The design of our relational environment is described by the hypergraph defining the universal relation. We present some or all of this structure to the user in graphical form to assist the user in discovering relationships among entities in the environment.

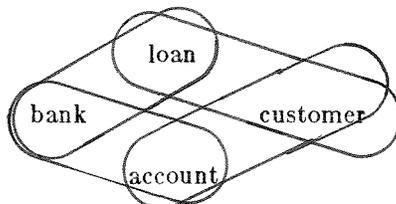

Figure 6: Hypergraph Showing the Objects for the Expanded Bank Database.

(2) *A graphical means of expressing queries* -- Although SQL (with our proposed extensions) suffices as a query language, we exploit the availability of our workstation to construct a simpler means of expressing queries.

### 4.2.1. Hypergraph Representation Theory

In order to understand the semantics of the universal relation from a formal standpoint, it suffices to treat the hypergraph abstractly. For a non-mathematically-oriented user to understand the semantics of the universal relation, it is helpful to have a graphical representation of the database. Such a graphical representation needs to be as clear and simple as possible. The issues that need to be addressed in constructing this interface involve both human-factors research and research in the theory of hypergraphs.

Hypergraph algorithms required to support the user interface include planarity and colorability algorithms. Just as planar graphs are easier to deal with than non-planar ones, so planar hypergraphs represent a clearer picture of the database than an equivalent non-planar representation. Of course, not all hypergraphs are planar, so we must concern ourselves with minimizing an appropriate measure of non-planarity. This measure can be the number of crossing hyperedges. Another useful measure is colorability. Objects (or maximal objects) can be represented in a particular color. A hypergraph representation is $k$-colorable if using $k$ colors, no intersecting hyperedges have the same color and no crossing hyperedges have the same color.

A user may focus on a particular subset of the set of all attributes. This may be because his view of the database excludes certain attributes, or it may be because the user wishes to restrict attention temporarily to a set of attributes relevant to the query being formulated. These cases require the system to compute an appropriate representation of the hypergraph in real-time, and thus poses an additional constraint on the design of hypergraph algorithms.

We note that many of the algorithms used in the system will be heuristic in nature, since finding optimal solutions may have exponential time complexity. For example, the hypergraph coloring problem stated above is *NP*-complete.

We are now in the process of developing a representation theory for hypergraphs and the construction of related algorithms. These algorithms will be analyzed for their asymptotic complexity as well as their expected-case performance in a real system. Since databases tend to be large (or to become large rapidly), asymptotic complexity is of practical as well as theoretical interest. Expected case behavior must be sufficiently good that a user will not lose his train of thought while waiting for the system to respond. It is hoped that we can design sufficiently fast algorithms that they can run on a personal workstation and not require the resources of a mainframe computer.

### 4.2.2. The Semantics of Graphical Queries

A primitive graphical query language is one in which the user may point at an object on the screeen and make a request of that object. Such a capability is useful to expand set-valued or relation-valued fields. Consider the root directory of Figure 2. By pointing to the "contents" field of the "book" tuple, the user could obtain a display of the book relation as shown in Figure 3. Similarly, the user may edit text by pointing to the contents field of a chapter tuple in the book relation. Thus, the semantics of the "expand" operation are context sensitive.

Although simple graphical queries like that described above go a long way towards simplifying the user's interaction with the system, our goals for a two-dimensional query language are much more ambitious. In a standard query language, the difficult queries are typically those that involve attributes that are related in several different ways (as in the expanded bank example of Figure 6).

We would like the user to use the hypergraph of the database as the basis for queries. The user selects those attributes of interest for the query. The system responds by showing the relationships among those attributes. The response either provides positive feedback to the user or prompts him to refine his query (e.g., the user may be interested only in the account relationship between banks and customers). In effect, the system is attempting to paraphrase the query to ensure that the user understands the semantics of the database.

The selection criteria that appear in the **where** clause are placed, in the graphical language, next to the attribute they relate to. Such graphical aids as rapid scrolling through a domain can serve to aid novice users and allow range queries to be specified without the use of the keyboard.

Selection criteria involving more than one attribute may be modeled by an arc connecting the relevant attributes and labeled by the comparison operator. Again the keyboard is not required. Updates, and the movement of data may be modeled by directed arcs.

Of course, these graphical statements may be augmented by statements entered from the keyboard using a variant of the System/U query language. Our goal, however, is to have a sufficiently powerful language that it is never *necessary* to resort to such a language.

The above remarks serve only to illustrate the "flavor" of our proposed language. The design of this language and the definition of its semantics are in a very early stage. In addition to the theoretical properties of the language, the human factors of this language require careful study.

## 5. Progress Report

The proposed project involves not only the development of a complete language for our operating system environment, but also its implementation. Since our research interests for this project involve the user environment, not the internals of the operating system, we are building our user interface on top of existing systems. Our specified plans are illustrated in Figure 7. We have available to us a VAX 780 running UNIX. On top of UNIX, we are running the relational database system ERIS. System/U (Korth, et al. [1984]), a prototype universal relation database system, runs on top of ERIS. These software components provide us with a foundation for our experimental research.

Using ERIS, we are building a relational operating system environment. This environment co-exists with the UNIX utilities, including the shell, so that traditional UNIX users will be able to communicate with users of the new environment. At this preliminary stage, the user interface is one-dimensional.

We are also currently developing the universal-relation-based graphical query system on top of System/U. This graphical interface is developed on top of the SUN personal workstation, which provides both graphic capabilities and local processing power.

There are still several issues that are currently being examined in regard to our ROSI project:

(1) *Directed graph organization* -- In a standard hierarchical file system, many files may have the same name, but no two files in the same subdirectory may have the same name. Thus, each file has a unique *pathname* which consists of the concatenation of the file-name with the names of all the directories one must traverse, starting at the root, to reach the file. Using *links*, it is possible to construct directories structured as arbitrary graphs.

Under our relational model, we may use pathnames for the same purpose. The interesting problem we are currently addressing is how to incorporate relations of relations into a relational language.

As a first attempt, we are extending SQL so that SQL statements may be applied recursively to relations within relations. Ultimately, we shall develop a language

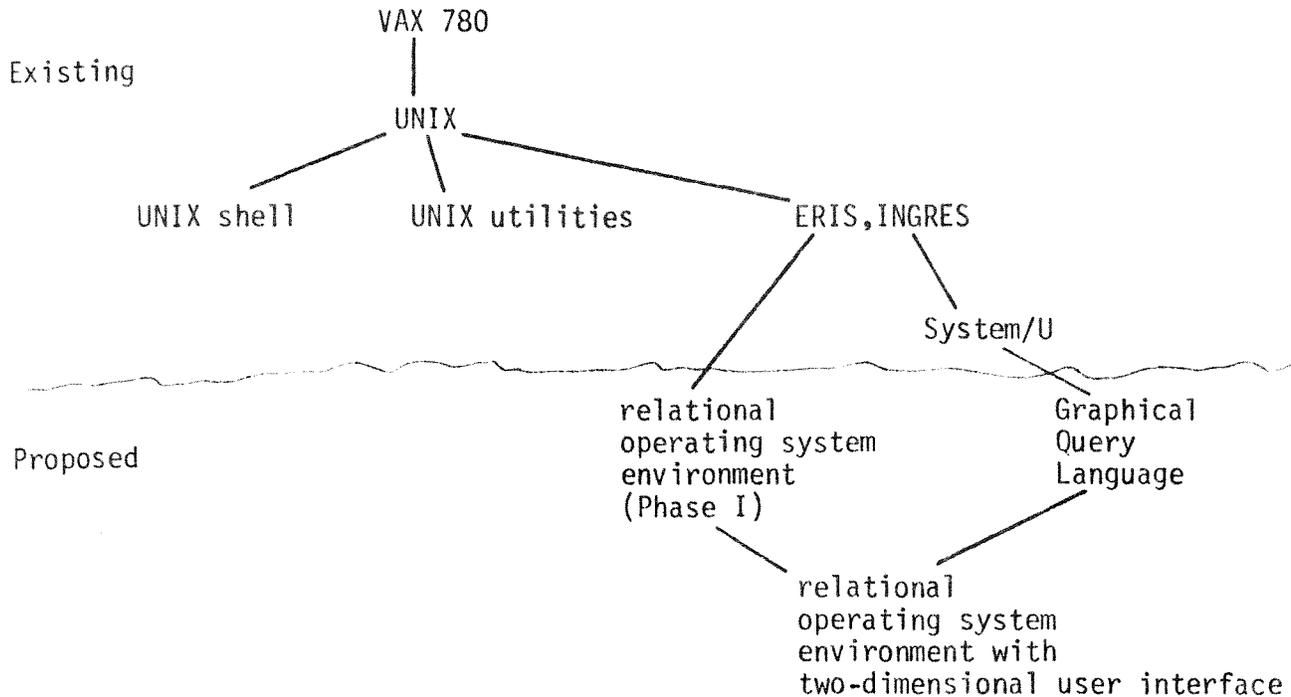

Figure 7: Existing and Proposed Environments

based on the ideas behind the universal relation model. Certainly, no user wants to see *the* universal relation, since it includes all directories, including those of other users. Rather, the user operates within a private universe consisting of personal and shared data. In extending the universal relation model and SQL to our directory structure, we must consider issues beyond the syntax of the query language. The semantics of insertion, update, and deletion need to be defined formally. The problems here are related to the view maintenance problem (Keller [1981]). A goal of this user view is to eliminate the need for user-maintained links. Furthermore, this view forms the foundation of our proposed graphical interface.

(2) *Non-first normal form structure* -- Relational theory assumes that the relations are at least in *first normal form* (1NF). That is, each attribute consists of a single atomic value which cannot be decomposed further. This assumption, however, cannot be satisfied easily in our approach. To illustrate this case, consider the mail system again. If one wishes to include a cc-list with the message, then the attribute cc-list cannot satisfy first normal form. Similarly, if one wishes to include an access-control-list (ACL) for protection purposes, the ACL attribute cannot be in first normal form. Finally, if one wishes to include in the relation chapter, the titles of all the sections, then again we have an attribute which cannot be in first normal form.

Jaeschke and Schek [1982] have investigated the relationship between 1NF and non-1NF relations with functional and multivalued dependencies. They have shown that under certain assumptions, multivalued dependencies on 1NF relations become functional dependencies in non-1NF relations. We shall need to extend their results to more general classes of data dependencies. Specifically, we must incorporate join dependencies into this theory because of their crucial role in the formal definition of the universal relation.

Furthermore, we propose to investigate efficient implementations of non-1NF relations.

To do this, we shall need to consider new normal forms that do not require first normal form. Although the user does not see a normalized database (the universal relation is rarely in normal form), a normalization theory for non-1NF databases is needed to allow us to implement our system efficiently. As is the case for standard normalization theory, we need to avoid the storage of redundant information and avoid the anomalies that can result from a poor relational database design.

We anticipate that existing normalization theory will provide a useful basis for this research despite the reliance of existing theory on first normal form.



\end{document}